\documentclass{aa}
\usepackage{graphicx}
\usepackage{txfonts}

\begin{document}

\title{Accretion Disc Evolution in DW Ursae Majoris: \\
A Photometric Study\thanks{based on observations obtained at Rozhen
National Astronomical Observatory, Bulgaria, at Hoher List, Germany and at 
Kryoneri, Greece}}

\author{
V. Stanishev\inst{1,2,}$^{\star\star}$ \and Z.
Kraicheva\inst{1,}$^{\star\star}$ \and H.M.J.
Boffin\inst{3,4,}$^{\star\star}$ \and  V. Genkov\inst{1,}$^{\star\star}$
\and C. Papadaki\inst{3,}$^{\star\star}$
\and S. Carpano\inst{5,}\thanks{
\email{vall@astro.bas.bg (VS), zk@astro.bas.bg (ZK),
henri.boffin@oma.be (HMJB), vgenkov@astro.bas.bg (VG), Christina.Papadaki@oma.be
 (CP), carpano@astro.uni-tuebingen.de (SC)}}}

\institute{Institute of Astronomy, Bulgarian Academy of Sciences,
           72, Tsarighradsko Shousse Blvd., 1784 Sofia, Bulgaria
   \and
           Present address: Physics department, Stockholm University, Sweden
   \and
    Royal Observatory of Belgium,
              Avenue Circulaire 3, B-1180 Brussels, Belgium
   \and
	European Southern Observatory, 
	Karl-Schwarzschild-Str. 2, 85748 Garching-bei-M\"unchen, Germany
   \and
	Institut f\"ur Astronomie und Astrophysik Universit\"at T\"ubingen, 
Abteilung Astronomie,  Sand 1, D-72076 T\"ubingen, Germany 
             }

\authorrunning{V. Stanishev et al.}
\titlerunning{Accretion Disc Evolution in DW Ursae Majoris}
\offprints{V. Stanishev} 

\date{Received ;accepted}

\abstract{We present an analysis of CCD photometric observations
of the eclipsing novalike cataclysmic variable \object{DW UMa}
obtained in two different luminosity states: high and
intermediate. The star presents eclipses with very different
depth: $\sim1.2$ mag in the high and $\sim3.4$ mag in the intermediate
state. Eclipse mapping reveals that this difference is almost
entirely due to the changes in the accretion disc radius: from
$\sim0.5R_{\rm L_1}$ in the intermediate state to $\sim0.75R_{\rm
L_1}$ in the high state ($R_{\rm L_1}$ is the distance from the white
dwarf to the first Lagrangian point). In the intermediate state, the
entire disc is eclipsed while in the high state, its outer part remains
visible.
We also find that the central intensity of the disc is nearly
the same in the two luminosity states and that it is the increase of the disc
radius that is responsible for the final rise from the 1999/2000 low
state. We find that the intensity profile of the disc is rather
flat and suggest a possible explanation. We also discuss the
effect of using a more realistic limb-darkening law on the disc
temperatures inferred from eclipse mapping experiments.
Periodogram analysis of the high state data reveals "positive
superhumps" with a period of $0\fd1455$ in 2002 and $0\fd1461$ in 2003, 
in accord with the results of Patterson et al. However, we cannot confirm the
quasi-periodic oscillations reported by these authors. We obtain
an updated orbital ephemeris of \object{DW UMa}: $T_{\rm
min}[HJD]=2446229.00687(9)+0\fd136606527(3)E$.
}

\maketitle

\section{Introduction}

\object{DW Ursae Majoris} is an eclipsing nova-like (NL) cataclysmic variable (CV) with
an orbital period of $\sim3.27$ hours. The star is one of the NLs which show
deep ($\sim4$ mag) low states (Hessman \cite{hess1}; Honeycutt et al.
\cite{hon, hon1}), and "negative" and "positive superhumps"
(Patterson et al. \cite{patt02}).
Shafter et al. (\cite{shaf1}) carried out the first detailed photometric and
spectroscopic study of \object{DW UMa} in a high state.
The photometry revealed $\sim1.5$ mag deep eclipses, with a peculiar V-like shape.
The spectral observations showed single-peaked H, \ion{He}{i} and \ion{He}{ii} emission
lines with superimposed transient narrow absorption components. These peculiarities
led Thorstensen et al. (\cite{th91}) to group \object{DW UMa} along with \object{SW Sex},
\object{V1315 Aql} and \object{PX And}, and to define so-called SW Sex stars.

Biro (\cite{biro}) presented $V$ and $R$ eclipse maps of \object{DW UMa} in a high state.
The eclipse maps show that the radial profile of the accretion disc (AD) temperature
distribution is rather flat, similarly to with what is found in other
\object{SW Sex} novalikes (Rutten et al. \cite{rutt}; Baptista et al. \cite{bsh}).
Biro (\cite{biro}) also estimated the distance to the system to be $270\,\pm50$ pc.

Spectral observations of Dhillon et al. (\cite{dhil}) in low state
revealed narrow Balmer emission lines originating from the
irradiated face of the secondary star. Marsh \& Dhillon
(\cite{mdh}) could not detect the secondary star in their $I$-band
spectra obtained in low state and derived a lower limit of the distance 450
pc or 850 pc depending on the assumed spectral type of the
secondary. Ultraviolet spectroscopy during a recent low state was
presented by Knigge et al. (\cite{knigge}). Surprisingly, they
found that during the low state the continuum shortward of 1450
\AA\ was higher and bluer than in the high state. This led Knigge et
al. (\cite{knigge}) to suggest that in the high state of \object{DW UMa}
the white dwarf (WD) is permanently hidden from our sight by the
rim of a flared accretion disc. Knigge et al. (\cite{knigge}) also modeled
the ultraviolet spectrum of the WD and estimated the distance to
\object{DW UMa} to be $830\,\pm150$ pc.

 \begin{figure*}[t]
 \includegraphics*[width=18cm]{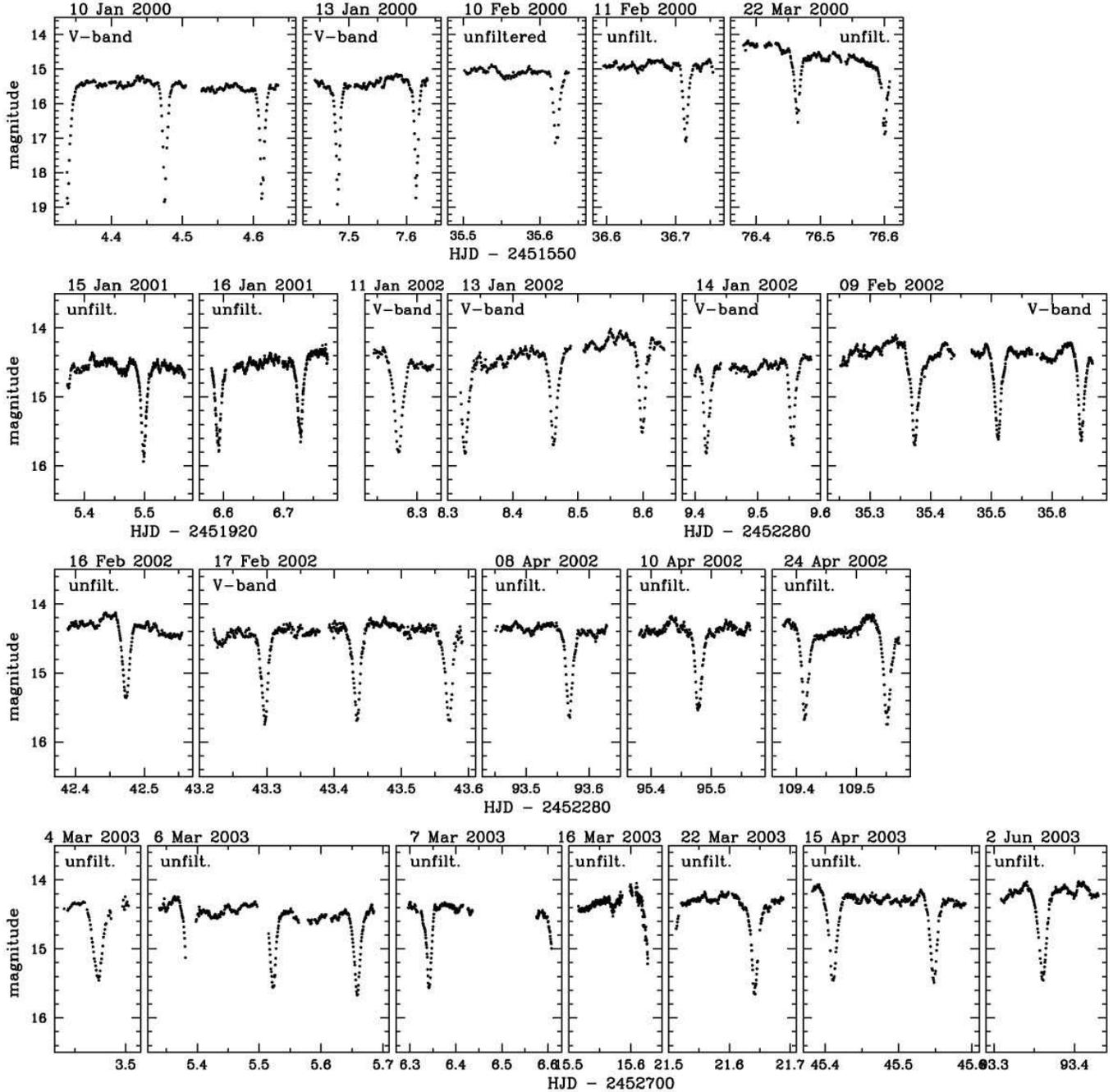}
 \caption{Photometric observations of \object{DW UMa}. Note the much larger
 range of the $y$-axis in the upper row panel.}
 \label{pats}
 \end{figure*}

Using the same observations as Knigge et al. (\cite{knigge}),
Araujo-Betancor et al. (\cite{syspar}) could measure the eclipse
phases of the WD. This allowed them to determine the system
parameters of \object{DW UMa} with very good accuracy. Repeating
the analysis of Knigge et al. (\cite{knigge}), and using $I$- and
$K$-band photometry, Araujo-Betancor et al. (\cite{syspar})
obtained two estimates of the distance to \object{DW UMa}: $930\,\pm150$
 pc and $590\,\pm100$ pc, respectively.

In this article we present  photometric observations of \object{DW UMa}
in two different brightness states -- intermediate and high.
We performed periodogram analysis in the high state and detected "positive
superhumps". We also present eclipse maps of \object{DW UMa}'s
accretion disc in the intermediate and high states. In the last part
of the article we discuss the results and present our conclusions.

\section{Observations and data reduction}

\begin{table}[t]
\centering 
\caption[]{CCD observations of \object{DW UMa}.}
\vspace{0.2cm}
\begin{tabular}{lccc}
\hline
\hline
\noalign{\smallskip}
 Date      & Start & duration  &  mid-eclipse  \\
                  &  HJD-2450000 & [hours] & [HJD]    \\

\hline
\noalign{\smallskip}
\multicolumn{4}{c}{Rozhen 2-m, Johnson $V$ filter }\\
\hline
\noalign{\smallskip}
2000 Jan 10  & 1554.34 & 7.13 & 1554.33889  \\
                          & & & 1554.47524  \\
                          & & & 1554.61264  \\
2000 Jan 13  & 1557.44 & 4.69 & 1557.48099  \\
                          & & & 1557.61743  \\
2002 Jan 11  & 2286.26 & 1.79 & 2286.27706  \\
2002 Jan 13  & 2288.32 & 7.50 & 2288.32641  \\
             &         &      & 2288.46296  \\
             &         &      & 2288.59911  \\
2002 Jan 14  & 2289.40 & 4.44 & 2289.41825  \\
             &         &      & 2289.55524  \\
2002 Feb 9   & 2315.25 & 9.95 & 2315.37403  \\
             &         &      & 2315.51110  \\
             &         &      & 2315.64799  \\
2002 Feb 17  & 2323.22 & 8.86 & 2323.29749  \\
             &         &      & 2323.43444  \\
             &         &      & 2323.57099  \\
\hline
\noalign{\smallskip}
\multicolumn{4}{c}{ROB Schmidt telescope, unfiltered }\\
\hline
\noalign{\smallskip}
2000 Mar 22  & 1626.38 & 5.50 & 1626.46748 \\
                          & & & 1626.60419 \\
2002 Feb 16  & 2322.39 & 3.96 & 2322.47774 \\
2002 Apr 8   & 2373.45 & 4.24 & 2373.56879 \\
2002 Apr 10  & 2375.38 & 4.43 & 2375.48054 \\
2002 Apr 24  & 2389.38 & 4.63 & 2389.41500 \\
                          & & & 2389.55209 \\
2003 Mar 16  & 2715.52 & 2.68 &            \\
2003 Mar 22  & 2721.51 & 4.23 & 2721.64166 \\
2003 Apr 15  & 2745.38 & 5.00 & 2745.41090 \\
                        &  &  & 2745.54838 \\
\hline
\noalign{\smallskip}
\multicolumn{4}{c}{Hoher List 1-m, unfiltered }\\
\hline
\noalign{\smallskip}
2000 Feb 10  & 1585.50 & 3.26 & 1585.62166 \\
2000 Feb 11  & 1586.60 & 3.80 & 1586.71482 \\
2001 Jan 15  & 1925.37 & 4.69 & 1925.49886 \\
2001 Jan 16  & 1926.58 & 4.64 & 1926.59172 \\
             &         &      & 1926.72804 \\
2003 Mar 04  & 2703.44 & 1.52 & 2703.47304 \\
2003 Mar 06  & 2705.34 & 8.31 & 2705.52281 \\
             &         &      & 2705.65868 \\
2003 Mar 07  & 2706.30 & 7.47 & 2706.34213 \\
\hline
\noalign{\smallskip}
\multicolumn{4}{c}{Kryoneri 1.2-m, unfiltered }\\
\hline
\noalign{\smallskip}
2003 Jun 02  & 2793.31 & 2.89 & 2793.36068 \\
\hline
\end{tabular}

\label{obs}
\end{table}

Photometric CCD observations of \object{DW UMa} were obtained with
the 2.0-m RCC telescope at Rozhen Observatory, the 0.85-m Schmidt
telescope at the Royal Observatory of Belgium (ROB), the 1-m
telescope at Hoher List Observatory and the 1.2-m telescope at 
Kryoneri Observatory between 2000 and 2003.
Depending on the telescope and the atmospheric conditions, the exposure
time used was between 20 and 60 s. A journal of the
observations is given in Table\,\ref{obs}. The CCD frames were
proceeded in the standard way with bias removal and flat-field
correction, followed by aperture photometry with the
 DAOPHOT procedures (Stetson \cite{ste}).
The Rozhen observations were performed with a Johnson $V$ filter,
while the rest are unfiltered. For Rozhen observations the stars DW UMa-3
($V$=16.0) and DW UMa-2 ($V$=17.043) from Henden \& Honeycutt (\cite{comp})
served as a comparison and check, respectively. For the unfiltered observations
as a comparison star we used the star \object{GSC 3822 0070}, whose standard
magnitudes are given by Biro (\cite{min1}). To be able to compare the $V$
photometry with the unfiltered one, we have corrected the unfiltered magnitudes
in the following way. During the observations in year 2002 \object{DW UMa} was
in a high state. Thus, we have computed the magnitude of
\object{GSC 3822 0070} in the instrumental system of the ROB Schmidt telescope
so that the average out-of-eclipse magnitude of \object{DW UMa} in that system
to be equal to the average out-of-eclipse $V$ magnitude computed from the
Rozhen data. We obtained $\sim13.02$ which is just between the $V=13.29$ and
$R=12.88$  magnitudes estimated by Biro (\cite{min1}).
We further assumed that \object{GSC 3822 0070} has the same magnitude
in the system of the Hoher List 1-m telescope and corrected those magnitudes
too. The Kryoneri photometry is relative to DW UMa-2. For DW UMa-2 we have
obtained an unfiltered magnitude of $\sim16.36$ and use it to put 
Kryoneri magnitudes in $V$. All the runs are shown in Fig.\,\ref{pats}.

 \begin{figure}[t]
 \includegraphics*[width=8.8cm]{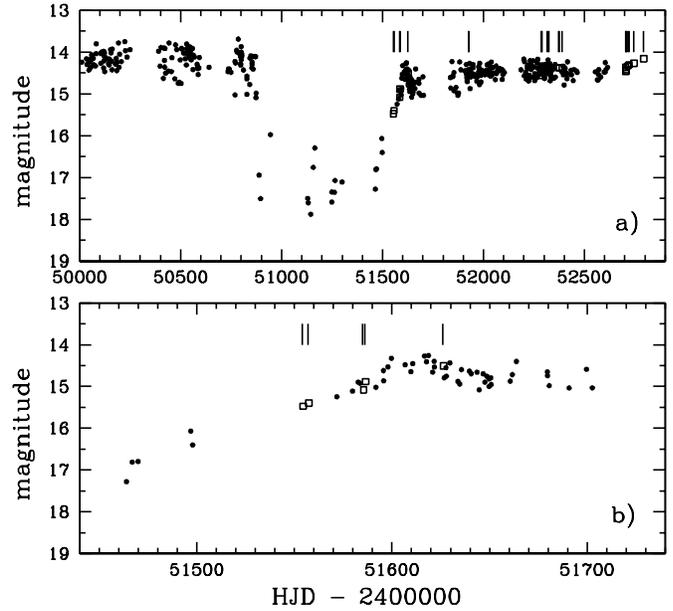}
 \caption{Part of the long-term light curve of \object{DW UMa}
 obtained by RoboScope
 (Honeycutt et al. \cite{hon1}) with the epochs of our observations
 marked with the vertical bars and the mean out-of-eclipse
 magnitudes shown with open squares. Our observations in 2000 have
 been obtained on the rising branch of a deep low state. For 
 clarity, the RoboScope data obtained during
 eclipse have been removed.}
 \label{lt}
 \end{figure}

From Fig.\,\ref{pats} one can see that in 2001-2003
 \object{DW UMa} was in its normal high state with
out-of-eclipse magnitude $V\simeq14.5$, but in 2000 we have caught
the star on the rising branch from a deep low state. The later is more
easily seen in Fig.\,\ref{lt} where we show part of the long-term
light curve of \object{DW UMa} obtained by RoboScope (Honeycutt et al. \cite{hon1}) with
the epochs of our observations marked with the vertical bars. In
Fig.\,\ref{lt} the open squares show the nightly mean
out-of-eclipse magnitude of \object{DW UMa} during our
observations. It is seen that our observations follow the
photometry of Honeycutt et al. (\cite{hon1}) very well. This also
gives us confidence in the correction of the unfiltered
observations described above.

\begin{figure}[t]
 \includegraphics*[width=8.8cm]{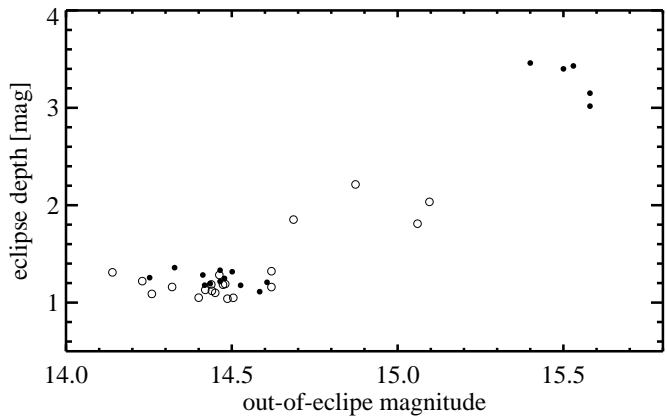}
 \caption{Eclipse depth as a function of the out-of-eclipse
 magnitude of \object{DW UMa}. Filled circles -- Rozhen $V$-band
 photometric observations; open circles -- unfiltered
 observations.}
 \label{depth}
 \end{figure}

Figure\,\ref{pats}  also reveals the huge difference in the
eclipse depth in 2001-2003 and Jan 2000: $\sim1.2$ and $\sim3.4$ mag,
respectively. To our knowledge the eclipses we see in 2000 are
among the deepest ever observed in a CV. The only close example is
\object{GS Pav}, which sometimes also shows $\sim3.2-3.3$ mag deep
eclipses (Groot et al. \cite{groot}). Figure\,\ref{depth} shows
the dependence of the eclipse depth on the magnitude at orbital
phase zero. The later was estimated by fitting low-order
polynomial to the out-of-eclipse measurements before and after the
given eclipse. The dependence is almost linear and the eclipse
depth decreases with system luminosity.

\section{Results}

\subsection{The orbital ephemeris}

The eclipse timings listed in Table \ref{obs} were determined by
fitting parabolas to the lower half of the eclipses. These timings
were analyzed together with those given by Dhillon et al.
(\cite{dhil}), Biro (\cite{biro}), Biro \& Borkovits  (\cite{min1}) and
Borkovits et al. (\cite{min2, min3, min4}) to obtain the
following orbital ephemeris:
\begin{equation}
T_{\rm min}[HJD]=2446229.00687(9)+0\fd136606527(3)E,
\end{equation}
that is very close to that derived by  Biro (\cite{biro}).
The {\it rms} around the fit is $\sim31$ s. We note that  Araujo-Betancor et al.
(\cite{syspar}) used a different ephemeris, which left them with (O-C)
residuals of +73 s. We note that our ephemeris (and that of Biro \cite{biro}) nearly
corrects for this residuals.

\subsection{Periodogram analysis}

After removing the measurements during the eclipses from the light
curves, the data obtained in 2002 and 2003 were searched for the "positive
superhumps" reported by Patterson et al. (\cite{patt02}). The
periodogram of the "raw" light curves did not give any conclusive
results. The most likely reason is the combination of very complex
light curves, removing $\sim1/3$ of the data around eclipse and
the sparse distribution of the observations. We would like to note
one peculiarity of the \object{DW UMa} light curves: in some of
the runs we clearly see a linear increase/decrease of the system
brightness (22 March 2000, 16 Jan 2001, 13 and 14 Jan 2002). The
same seems to be also present in the observations of Biro
(\cite{biro}; see his Fig.\,1) too.
This  suggests that \object{DW UMa}'s mean magnitude varies on time scale of
$\sim1-2$ days. We note that a 2.2 days (in 2002) or 2.1 days (in 2003) periodicity could be linked
to the superhump we have detected in the light curves. It is however still premature to 
claim that such a periodicity as been found.
From our Figs.\,\ref{pats} and \ref{depth}, and  
Biro's Fig.\,1,
one can see that these brightness variations are not
accompanied by variations of the eclipse depth. This suggests a
global brightening of the AD without significant changes of its
brightness distribution or radius.

\begin{figure*}[t]
 \includegraphics*[width=18cm]{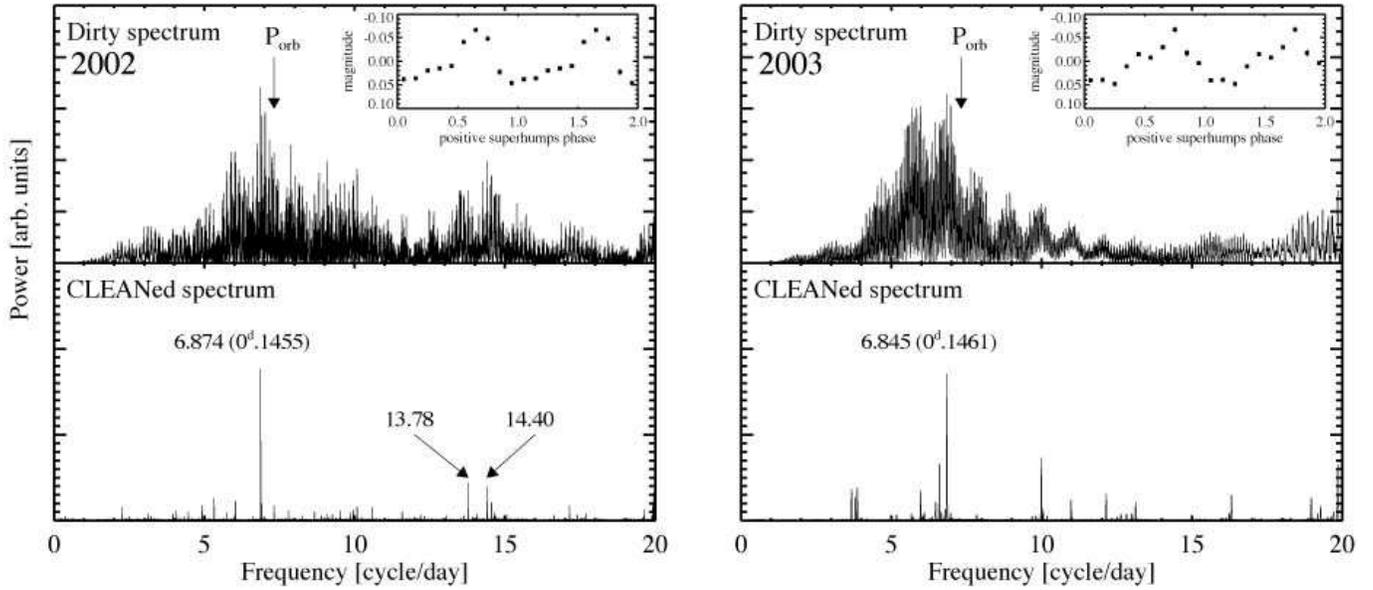}
 \caption{Periododgram analysis of \object{DW UMa} in 2002 -- {\it left} and in
 2003 -- {\it right}. The insets show the mean superhump shape.}
 \label{per}
 \end{figure*}

To account for these trends we fitted straight line to all the runs, subtracted it
from them and repeated the periodogram analysis. To de-convolve the computed
power spectrum with the spectral window, the CLEAN algorithm of Roberts et al.
(\cite{rob}) has been used. The results are presented in Fig.\,\ref{per} and show
that in 2002 data we could detect "positive superhumps" with a period of $P_{\rm sh}^+=0\fd1455$,
which is very close to that of  Patterson et al. (\cite{patt02}).
We also see two of the other three peaks reported by  Patterson et al. (\cite{patt02}),
those close to the
first harmonics of the orbital and "positive superhumps" periods. The multi-sinusoidal
fit to the data gives the following full amplitudes of these modulations:
0.096, 0.053 and 0.044 mag, in order of decreasing period.
In 2003 we also detect "positive superhumps" with a period of 
$P_{\rm sh}^+=0\fd1461$ but none of the shorter signals.  The full amplitude of the
modulation is $\sim0.08$ mag.

\begin{figure}[!h]
 \includegraphics*[width=8.8cm]{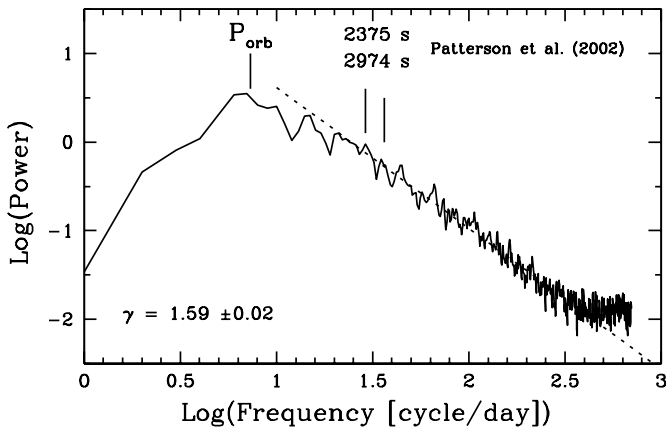}
 \caption{Mean power spectrum of \object{DW UMa} in double logarithmic scale.}
 \label{ps}
 \end{figure}

We also searched for periodic and quasi-periodic oscillations (QPOs)
on the minute time scale. The presence of
QPOs with characteristic periods in the range 1000--4000 s in NLs is
now a well established observational fact (for a short review see
Patterson et al. \cite{patt02}). Patterson et al. (\cite{patt02}) also
see in a few \object{SW Sex} stars, including \object{DW UMa},
stable oscillations with periods in the same range as the QPOs.
Figure\,\ref{ps} shows the mean power spectrum of \object{DW UMa} runs
in double logarithmic scale. The power spectrum has  a typical "red noise"
shape with a power-law decrease of the power with frequency
$P(f)\propto f^{-1.59 \pm0.02}$ and shows no
traces of QPOs. We also do not see the stable oscillations with periods
of 2375 s and 2974 s reported by Patterson et al. (\cite{patt02}).
The "red noise" shape of the power spectrum is believed to be a manifestation of
the flickering, which is strong in \object{DW UMa}  and as usual appears
as well defined peaks with a typical duration of 5--15 min, and amplitudes
reaching $\sim0.2-0.3$ mag. The standard deviation of the light curves
after subtraction of the best fit is $\sim0.06$ mag, which is consistent with
what is seen in other NLs (Kraicheva et al. \cite{krtt2, krmv}; Stanishev
et al. \cite{stpx}).

We also note that in some of the runs, 2 Feb 2002 and 24 Apr 2002 for example, 
a hump just before the eclipse is clearly seen. Such humps are commonly observed 
in high-inclination dwarf novae in quiescence and are caused by the hot spot formed
where gas stream hits the outer edge of the disc. The spot also makes the eclipse 
to appear asymmetric, with a well pronounced shoulder on the egress. The eclipses of 
\object{DW UMa}, including these with preceding humps are fairly symmetric and with no 
hints of hot spot. We believe that the "pre-eclipse humps" seen in \object{DW UMa}
are a result of superhumps in conjunction with strong flickering activity.

\subsection{Eclipse mapping}

The eclipse mapping method (Horne \cite{hor}) and the Maximum Entropy
technique (Skilling \& Bryan \cite{ski}) was used to recover the
surface brightness distribution of the accretion disc in
\object{DW UMa}. The code we use (written by one of us -- V.S.)
allows for the presence of light that is never eclipsed. We assume
a flat disc which lays in the orbital plane. Eclipse mapping
assumes that all brightness variations result from the eclipse of
the disc by the secondary star. Thus, any brightness variations
due to anisotropic emission from the hot spot or superhumps cannot
be handled by the algorithm. To account for these brightness
variations we fitted the out-of-eclipse flux with a low-order
polynomial function and normalized the eclipse by the fit.

\begin{figure*}[t]
 \includegraphics*[width=18cm]{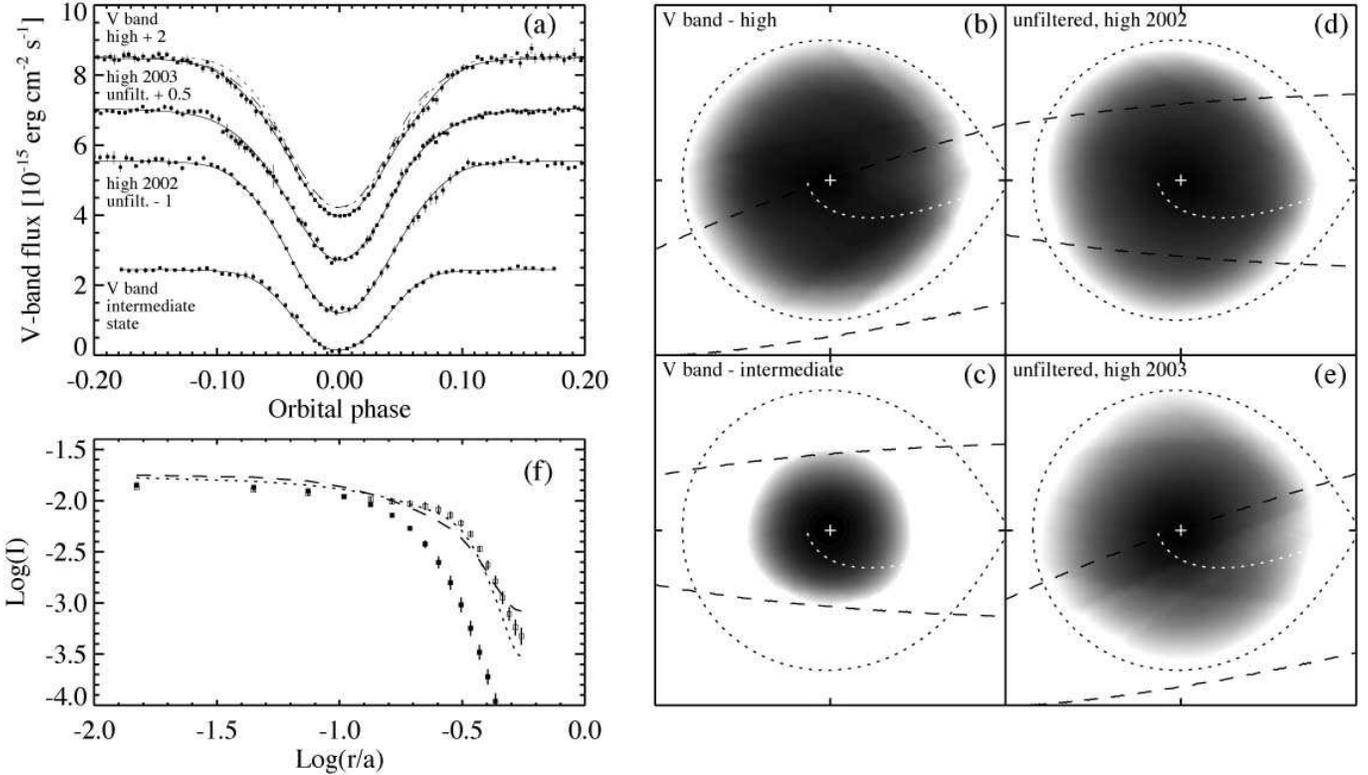}
 \caption{Eclipse mapping of \object{DW UMa}. {\bf (a)}
  mean light curves with the best fits. The dotted and the dashed lines are 
 the fits to the unfiltered
  high state eclipses in 2002 and 2003, respectively, overplotted on the $V$-band high state one;
 {\bf (b-e)} accretion disc eclipse maps, the WD's Roche lobe is shown with dotted
  line and the shadow of the secondary at orbital phases 0.0 and 0.041 with
  dashed line; {\bf (f)} radial intensity profiles
 of the accretion disc: open squares -- $V$-band high state,
  filled squares -- $V$-band intermediate state, and dotted and dashed lines 
  --  unfiltered high state in 2002 and 2003, respectively.}
 \label{fit}
 \end{figure*}

The shape of the individual eclipses is highly variable, in particular in the upper
half of the profile. This is most likely due to the flickering and is not related
to real changes of the AD structure.
As mentioned before, flickering in \object{DW UMa} is strong and certainly
can affect the upper part of the eclipse profile.
To reduce the influence of flickering and other noise
the eclipses were averaged in small phases bins. 
We note that even if some of the "pre-eclipse humps" discussed above
are caused by temporary appearance of a hot spot and in these cases the eclipse shape
variations are not caused by flickering, when working with an average eclipse the 
spot will not be detected in the eclipse maps.
Because of the large difference in
the eclipse depth we have analysed the eclipses
in the high and intermediate state separately. In addition,
the $V$-band and the unfiltered high state eclipses were also
analysed separately. In the high state we have enough
eclipses to try to reduce the influence of the flickering even more:
we  averaged only the lowest 70\% of the point in each bin. Such an
approach  is reasonable because
flickering increases the observed flux. The error bars assigned
to the mean values are the standard errors of the mean.
The mean eclipses are shown in Fig.\,\ref{fit}$a$. They 
have been scaled so as to have out-of-eclipse flux equal to the
mean flux at orbital phase zero. The mean fluxes at orbital phase zero
were determined from the fits used to normalize the eclipses.

To use the eclipse mapping method one needs to know the orbital inclination
$i$ and the mass ratio $q=M_2/M_1$, where $M_1$ is the mass of the
WD and $M_2$ is the mass of the secondary star. In an eclipsing CV,
$i$ and $q$ are related trough the duration of the eclipse of the
WD $\Delta \phi=\phi_e-\phi_i$, where $\phi_i$ and $\phi_e$ are
the ingress and egress phases of the WD's center. In NLs, $\phi_i$
and $\phi_e$ can only be directly measured in the low state.
Araujo-Betancor et al. (\cite{syspar}) obtained UV observations of
\object{DW UMa} in the low state and could measure
$\phi_i$ and $\phi_e$. This allowed them to tightly constrain the
system parameters. More specifically they obtained
$q=0.39\,\pm0.12$ and $i=82^\circ\,\pm4^\circ$, and we used these
values in our eclipse mapping analysis\footnote{The detection of
"positive superhumps" in \object{DW UMa} indicates a lower value
of $q\sim0.30-0.33$. This is well in the $1\sigma$ range given by
Araujo-Betancor et al. (\cite{syspar}). We tried eclipse mapping
with $q=0.3$, but this did not change the results significantly.
The results presented below are obtained with $q=0.39$}.

In certain circumstances $\Delta \phi$ can be also measured from high
state eclipses (Wood et al. \cite{wood92}). Such measurements rely on
the fact that in certain combinations of $i$ and $q$, at $\phi_i$ and
$\phi_e$ approximately half of the disc area is eclipsed.
This is the case for most deeply eclipsing systems, such as \object{DW UMa},
provided that $q\geq0.3$ and $i\geq80^\circ$. Thus,
$\Delta \phi$ can be measured from the points of
{\it half-intensity} of the eclipse. In almost all  NLs
this turns out to be the only method available to measure $\Delta \phi$.
However, the accuracy of these measurements is usually very low, and
\object{DW UMa} itself is a good example. Published values of $\Delta \phi$
measured from high state eclipses range from $\sim0.073$
(Shafter et al. \cite{shaf1}) to  $\sim0.090$ (Dhillon et al.\cite{dhil};
Biro \cite{biro}). From our high state eclipses we determine
$\Delta\phi\sim0.064-0.068$, the lowest values ever measured in \object{DW UMa}.
The only measurements of $\Delta \phi$ close to the accurate value of
Araujo-Betancor et al. (\cite{syspar}) have been obtained when the AD
contribution is small, $\sim0.083$,  by Dhillon et al. (\cite{dhil}) in the low
state. From the intermediate state eclipses we obtained
$\Delta\phi\simeq0.080$, which is also very close to the value of
Araujo-Betancor et al. (\cite{syspar}). It is clear then that measurements of
$\Delta \phi$ from eclipses in the high state and system parameters
determinations based on these have to be interpreted with
extreme caution.

The eclipses are rather symmetric, which  suggests a nearly symmetric
distribution of the AD intensity. That is why in
the reconstructions we used a "full azimuthal streaming" default image
(Horne \cite{hor}). The eclipses were fitted to reduced $\chi^2=1$ and
the fits are shown with solid lines in Fig.\,\ref{fit}$a$. The corresponding eclipse
maps are shown in logarithmic gray-scale in
Figs.\,\ref{fit}$b-e$. The outermost disc parts displayed
correspond to the level where the intensity drops below 10\% of the maximum,
which we adopt as measure of the disc radius.
The corresponding radial profiles of the disc images
are shown in Fig.\,\ref{fit}$f$. The main results of the eclipse mapping of
\object{DW UMa} can be summarized as follows:

\begin{itemize}
\item the AD radius increased from $\sim0.5R_{\rm L_1}$ in the
intermediate state to $\sim0.75R_{\rm L_1}$ in the high state ($R_{\rm
L_1}$ is the distance from the WD to the first Lagrangian point
$L_1$);

\item the AD brightness distribution in both states is very flat;

\item the central intensities of the high and intermediate state maps
are very close (Fig.\,\ref{fit}$f$). Thus,  it seems that the
increase of the AD radius can fully account for the long-term
brightness variations, at least for the last one magnitude on the
rising part of the 1999 low state.

\item in the high state a significant amount of uneclipsed light is detected:
$\sim9$\%  in the $V$-band and $\sim17$\% in the unfiltered maps.
The uneclipsed light is essentially zero in the intermediate state.

\item in the intermediate state the AD is almost totally eclipsed, whereas in the high state,
a large part of the disc remains visible at mid-eclipse. This is the reason for the
large difference in the eclipse depth.

\end{itemize}

\section{Discussion}

Perhaps the most intriguing results of this study are the AD radius variations 
and the nearly equal central intensities of the high and intermediate 
state maps. As high and low states are caused by strong variations of mass 
accretion rate, the later is difficult to explain within the 
standard accretion disc theory. However, what we formally call "intermediate 
state" is in fact a transition from the 1999/2000 low state to the high state 
and during this transition the disc might not be in equilibrium with the actual 
accretion rate. However, this is most likely not the case. Figure\,\ref{lt}b shows 
that the recovery from the low state is rather slow and takes $\sim4$ months, 
i.e. the disc has enough time to adjust its structure to changes of mass 
accretion rate. This suggests that at given value of the accretion rate some 
kind of saturation occurs and further increase of the accretion rate does not 
lead to increase of the emission from the inner part of the disc. Instead, 
the rise of the system luminosity is due to an increase of the disc 
radius from $\sim0.5R_{\rm L_1}$ to $\sim0.75R_{\rm L_1}$\footnote{We stress that the fact that we see 
this behavior on the rising branch of the low state does not necessarily mean that 
one should expect the opposite on the falling branch.}.
These values are in accord with the AD radii in NLs, including \object{DW UMa} itself,
determined by Harrop-Allin \& Warner (\cite{haw}) from high- and low-state observations.
Note that the ratio of the area of circles with these radii is $\sim2.25$. If the AD 
brightness distribution does not depend on radius, this will give $\sim1$ mag increase 
of the brightness, as observed.
The reason of this behavior is not clear.
It could be that at certain value of the accretion rate some mechanism starts 
removing energy from the inner disc, thus preventing further increase of the 
emission of this part. Thus, it appears that this issue could be tightly related 
to the mechanism responsible for the flat intensity profile of the disc, 
which we discuss next.

The flat AD intensity profile found in \object{SW Sex} type NLs is
still a puzzle. This problem has been extensively discussed in the
literature, but there is no widely accepted model. Currently,
two of the proposed models seems to be most viable.
Accretion disc wind (or something else) can remove energy from the
inner part of the disc thus causing the profile to be flatter than
the steady-state $T_{\rm eff}\propto r^{-3/4}$ dependence. The second
model assumes that \object{SW Sex} stars are intermediate polars
and that the inner part of their discs is missing. 
In order to account for the other properties of the \object{SW Sex}  
stars it is also assumed that after the first impact with the AD's edge,
part of the gas in the accretion stream continues moving above the 
disc and hits its surface again close to the WD. 
In the intermediate polar model the stream hits the WD's magnetosphere
(for a review see Hellier \cite{helrev}).
In this later model the flat
profiles found in the eclipse mapping experiments are simply a
result of the assumption that the AD extends down to the WD's
surface. The later model has recently gained additional support.
Rodriguez-Gill et al. (\cite{gil1, gil2}) reported the discovery
of periodic circular polarization in two \object{SW Sex} stars.
Patterson et al. (\cite{patt02}) reported possible detections of
periodic signals ($\sim1000$ s) in the light curves of several
\object{SW Sex} and suggested that this could be a manifestation
of the rotation of a highly magnetic WD. Note, however, that in the 
intermediate polars the radius of the WD's magnetosphere or equivalently
the inner radius of the AD $r_{\rm in}$ depends on the mass accretion rate
$\dot{M}$: $r_{\rm in}\propto\dot{M}^{-2/7}$. Thus, the higher $\dot{M}$ 
the lower $r_{\rm in}$. Therefore, it can be expected that in the high-state
eclipse mapping analysis the AD brightness distribution will appear 
closer to the steady-state law than that in intermediate state.
We do not see this behavior and the AD brightness distribution is rather
flat in both state eclipse maps.

Recently, Knigge et al. (\cite{knigge}) found that during the 1999
low state, the UV continuum of \object{DW UMa} shortward of 1450 \AA\ was
higher and bluer than in the high state.  The authors suggest that in 
the high state the WD and the inner hot part of the disc are permanently 
hidden from our sight by the rim of a flared AD.
This certainly can alter the disc intensity profile of \object{DW
UMa} and cause it to appear flat in the eclipse maps. However, the
weakness of this model is the need of a very high AD
rim; its height should be at least $\sim0.15R_{\rm disc}$ if
the system inclination is $82^\circ$. With the present knowledge, it is 
difficult to understand how such a high rim can be formed. Moreover, it 
should have a constant height because Knigge et al. (\cite{knigge}) found that in
the low state the UV spectrum is {\it at all} orbital phases bluer and the flux higher
than in high state. In the classical steady-state AD model this would
require unrealistically high mass accretion rate. The impact of
the accretion stream with the disc can form a thick bulge but it
could not persist along the whole disc rim.

Here, we suggest another possibility. Bobinger et al. (\cite{bob})
found that the intensity profile of \object{IP Peg}'s AD in
decline from outburst is flat. The authors explored many possible
causes of this and concluded that the presence of an optically
thick and geometrically flat layer on top of the AD's surface
masks its true intensity profile. Bobinger et al. (\cite{bob})
suggested that this layer may be formed by the wind from the disc.
This is even more relevant in NLs, which generally have hotter
discs than dwarf novae.

Knigge \& Drew (\cite{knigge97}) modeled the wind in the NL
\object{UX UMa} and found that between the disc's photosphere and
fast-moving part of the wind there is a transition zone of a slowly outflowing
and relatively dense layer. We suggest that such a layer may
also be present in \object{DW UMa}, and may shade the WD and the
inner hot part of the disc, or what is more likely, attenuate
the radiation of the inner disc and the WD. Although this may seem somewhat
speculative, it can account for (1) the flat disc profile; (2) the
red UV spectrum in high state and (3) nearly the same temperature of
the inner disc in the high and intermediate states. However, it
should be borne in mind that in \object{UX UMa} itself this layer
does not flatten the AD's temperature profile; \object{UX UMa} is
often given as an example where the results of eclipse mapping are
consistent with the steady-state $T_{\rm eff}\propto r^{-3/4}$
law. The reason for the difference may be different parameters of
the wind.

 \begin{figure*}[t]
 \includegraphics*[width=12cm]{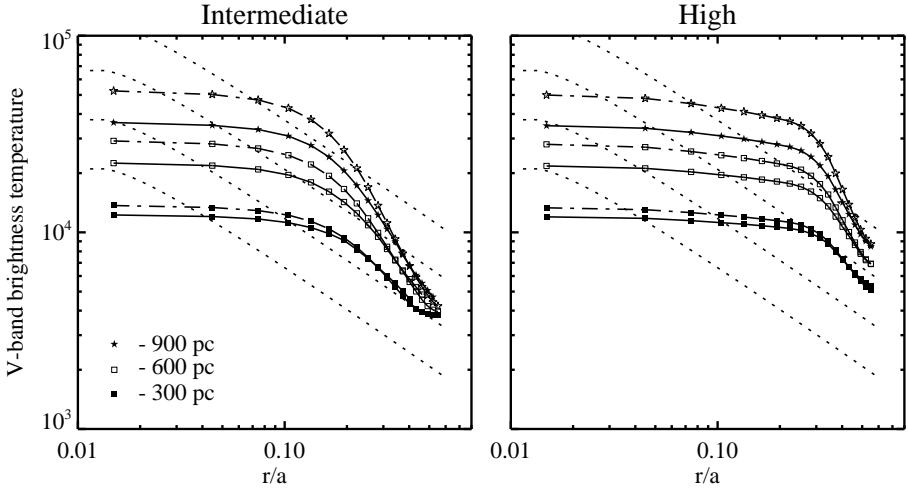}
 \caption{Radial profile of the AD brightness temperature
 obtained  by using (i) temperature dependent limb-darkening coefficient
 -- symbols connected with solid lines
 and (ii) with constant limb-darkening coefficient of 0.6 --
 symbols connected with dash-dotted lines.
 The dotted curves show the effective temperature of steady-state
 ADs for mass transfer rates
 $\dot{M}=10^{-10}-10^{-7}\,M_{\sun}$\,yr$^{-1}$ onto 0.77$M_{\sun}$ WD.}
 \label{rad}
 \end{figure*}

Having the eclipse map calculated one can estimate the
brightness temperature $T_{\rm BR}$ of the accretion disc
assuming blackbody emission.
This is done for the $V$-band eclipse maps 
by solving for $T_{\rm BR}$ the following equation:
\begin{equation}
f_j=\frac{C\delta_j\cos i}{D^2}\,\frac{\int v(\lambda)B_\lambda(T_{BR,j})\lambda\,d\lambda}
{\int v(\lambda)\lambda\,d\lambda}.
\label{temp}
\end{equation}
Here $f_j$ is the flux emitted by the $j$-th element of the AD,
$T_{\rm BR,j}$ and $\delta_j$ are its brightness temperature and
area, respectively, $B_\lambda(T_{\rm BR,j})$ is the blackbody
spectral distribution, $v(\lambda)$ is the response of the $V$
bandpass and $D$ is the distance to the system. As noted by
Baptista et al. (\cite{bap98}) the relation between the brightness
temperature and the effective temperature is not trivial and to
obtain such a relation the vertical structure of the disc has to
be known. However, the emission of the hot optically thick ADs in
novalikes probably closely follows the blackbody law  and $T_{\rm
BR}$ should be close to the effective temperature. Optically thick
ADs are however subject to an effect analogous to the
limb-darkening in stars and the factor $C$ takes this into
account. This later effect greatly complicates the situation
because it is well known that 1) the real limb-darkening differs
significantly from often used linear law and 2) $C$ is a function
of surface gravity, temperature and wavelength (Wade \&
Rucinski \cite{ld}; Diaz et al. \cite{diaz}).

To investigate the importance of limb-darkening further, we have
calculated $C$ by using the available synthetic stellar spectra
with solar abundances and $\log g=5$ calculated by R.L.
Kurucz\footnote{available at {\tt http://cfaku5.harvard.edu}}. We
calculated $C$ for $V$-band wavelengths, $i=82^\circ$ and
temperatures from 5000 to 50\,000\,K. Not surprisingly $C$ turns
out to depend strongly on the temperature (non-linearly) and
changes from $\sim0.35$ at $T_{\rm eff}=5000\,K$ to $\sim0.84$ at
$T_{\rm eff}=50\,000\,K$. Thus, it seems that the dependence of
limb-darkening on the temperature can affect significantly the
estimation of AD brightness temperature obtained from eclipse
mapping. This dependence would also affect the obtained radial
profile of the temperature in ADs with {\it temperature varying in
a wide range}. This would have a direct consequence on the
inferred accretion rates and distances to the CVs studied by
eclipse mapping.

Although ADs structure is different from that of the normal stars, Diaz et al. (\cite{diaz})
have shown that the difference between limb-darkening laws obtained with Kurucz stellar
spectra and AD
model calculations is small (see Fig.\,1 in Diaz et al. \cite{diaz}).
Thus, the main source of error when calculating $C$ is the assumption of
constant surface gravity $\log g=5$ ($\log g=5$ is the maximal
gravity in Kurucz' models). In real discs the gravity changes from
$\log g\sim4.5$ at the outer edge to $\log g\sim6.5-7.0$ at the inner parts.
The work of Wade \& Rucinski (\cite{ld}) shows that in stars for given effective temperature
the limb-darkening coefficient decreases with increasing $\log g$, which means
$C$ increases with $\log g$. If this holds true for ADs and $\log g>5$, 
then the temperatures one
obtains with $\log g=5$ overestimate the real temperature of the AD, particularly in its
inner part, where $\log g>5$; the degree of overestimation is however difficult to
estimate without model spectra with $\log g>5$.

Since the distance to \object{DW UMa} is rather uncertain
(see the Introduction for details) we have solved Eq.\,\ref{temp}
for three distances: 300, 600 and 900 pc. Figure\,\ref{rad} shows the
results: symbols connected with solid lines
show solutions with limb-darkening
as determined from Kurucz' model spectra, and those connected with dash-dotted
lines show the brightness temperature obtained using a linear
limb-darkening law with a coefficient of 0.6.
 Also shown with dotted curves is the expected
effective temperature radial distribution of steady-state ADs for
mass transfer rates $\dot{M}=10^{-10}-10^{-7}\,M_{\sun}$\,yr$^{-1}$
onto $0.77M_{\sun}$ WD (Araujo-Betancor et al. \cite{syspar}). It
is clearly seen that the temperature of the AD in \object{DW UMa}
in both high and intermediate state is very far from following the
steady-state radial dependence: it is much flatter that expected.
This finding is in accord with the results of Biro (\cite{biro}).
Also seen is that the usage of more realistic limb-darkening law
results in lower AD temperatures. The effect on the radial
dependence is however rather small, which is a direct consequence
from the flat intensity profile.

 \begin{figure*}[t]
 \includegraphics*[width=18cm]{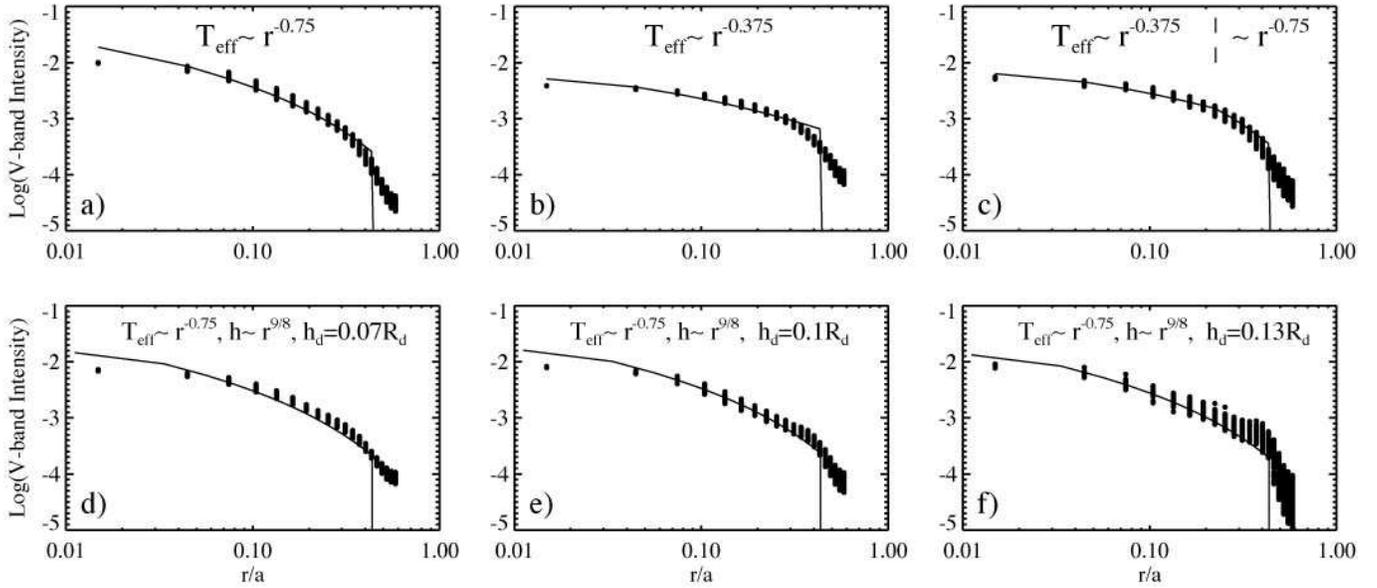}
 \caption{Reconstruction of the radial $V$-band intensity profile of simulated eclipses with
  S/N ratio of 100. Dots -- the reconstruction, solid line -- the disc profile used to
  simulate the eclipse. {\bf a--c} -- flat discs with different radial dependence of the 
  temperature, {\bf d--f} 3D steady-state discs with emitting outer rims of different height 
  (see text for details). In all reconstructions the eclipses were fitted to reduced $\chi^2=1$ 
  using a flat 1D disc geometry and full azimuthal streaming default images.}
 \label{simul}
 \end{figure*}

Although the recent studies point to a rather large distance to
\object{DW UMa}, $\geq500$ pc, we note that this seems
inconsistent with the results of eclipse mapping (Biro
\cite{biro}; this study). The main source of concern is the
relatively high temperature ($\sim20\,000$\,K) at the outer disc
regions obtained when distance larger than 500 pc is assumed. To
achieve this, a very high mass accretion rate would be needed. It
seems not very likely that the reason for this is the usage of
a simplified limb-darkening law. The cause of the discrepancy is a
puzzle. Although the dependence of limb-darkening on both the
temperature and surface gravity is in favor of obtaining flatter
radial temperature profiles (compared to if constant
limb-darkening coefficient is used), this could hardly explain the
very flat profile in \object{DW UMa}. The tests of the algorithm
over simulated eclipses showed a good
recovery of the radial profile.
 Results of some of the simulations of flat discs are shown 
in Figs.\,\ref{simul}{\it a-c} 
-- {\it (a)}  the $V$-band intensity profile of a steady-state accretion disc with 
$T_{\rm eff}\propto r^{-0.75}$;  
{\it (b)} $T_{\rm eff}\propto r^{-0.375}$ and {\it (c)} a combination of both. 
The disc elements are assumed to emits as a black body at given temperature. 
The emergent radiation is passed through a $V$-band filter, foreshortened and limb-darkened. 
We used a linear limb-darkening law with a coefficient 0.6. Gaussian noise of S/N=100 
was added and so simulated eclipses were fitted to reduced $\chi^2=1$ using a full azimuthal streaming 
default image. It is seen that 
the recovery in the middle part of the disc is good. Figure\,\ref{simul} 
also clearly shows one well-known problem of the ME technique, namely that it tends to 
smear out any sharp structures in the disc. As a result the outer edge of the disc could not be
accurately reconstructed and the radial profile is steeper at the outer part.
On the other hand, in the inner disc only the very
innermost pixels are smoothed and this is not a serious 
problem for the recovery of the radial profile. We notice that decreasing the noise level to
S/N=300 does not help much to obtain a better reconstruction and the above situation does not 
improve significantly.

Of more serious concern are the various systematic effects that may become 
important in the nearly edge-on systems. As already discussed, the mass accretion rate 
in NLs is probably not high enough for the disc rim to obscure the disc surface if 
the inclination is $\sim80^\circ$.
However, if $\dot{M}\sim10^{-8}\,M_{\sun}$\,yr$^{-1}$ the disc rim may still be 
as high as $\sim0.07-0.1R_{\rm disc}$\footnote{This refers to the height above the disc mid-plane.
The actual size of the rim is twice larger.}.
Recently, Smak (\cite{smak}) suggested that in this case if the temperature of the rim is 
comparable to that of the outer disc, in
the high-inclination NLs the rim may contribute more than 50\% of the 
system light. Clearly, this would affect
the eclipse profile significantly and when not accounted for may bias 
the eclipse mapping results.

Smak (\cite{smak94}) and Bobinger et al. (\cite{bob})
have studied some aspects of the reliability of the reconstruction of 3D discs.
Of particular interest is the work of Smak (\cite{smak94}) who tested
the reliability of the reconstruction of 3D discs if 
a simple flat 1D disc geometry is used in the reconstruction. 
Smak's work is however not based on the ME technique.
As a part of the testing of our eclipse mapping code we have 
performed a similar study. We simulated eclipses in a binary with
 $i=82^\circ$, $q=0.39$, $M_1=0.77M_\odot$ and $\dot{M}\sim10^{-8}\,M_{\sun}$\,yr$^{-1}$. 
A steady-state disc, whose (half)thickness varies as $r^{9/8}$ is assumed.
The outer disc rim is modeled as a cylindrical surface with a temperature equal to 
that of the outermost part of the disc. The disc has a radius of $0.75R_{\rm L_1}$. 
We then simulated eclipses for three values of
the rim height, 0.07, 0.1 and $0.13R_{\rm disc}$, which are appropriate for very high
mass accretion rates. 
The $V$-band intensities were computed as in the case of the flat disc 
simulations\footnote{Note that in the disc geometry assumed, the angle at  which  given disc element is seen is 
also a function of the orbital phase. In this case the foreshortening and limb-darkening
have to be applied simultaneously and they become part of the visibility function 
(or projection matrix) applied to the disc elements.}. Gaussian noise of S/N=100 
was added and the simulated eclipses were fitted assuming a flat disc geometry.

The reconstructed radial $V$-band intensity profiles are shown in Figs.\,\ref{simul}{\it d-f}.
Generally, our results show that the radial 
profile of a 3D accretion disc may be still well reconstructed assuming a flat geometry, 
provided that the inclination is not very high.
However, the reconstruction clearly worsens with increasing 
the rim height and in Figs.\,\ref{simul}{\it d-f} one can see the increasing contribution 
of the rim emission. Thus, for very high-inclination NLs with $i>85^\circ$ this will definitely 
be a problem, especially when the rim stars shielding part of the disc. 
Note that the impression for a better reconstruction of the disc's edge,
compared to the flat disc simulations is misleading. This is due to the additional light from
the rim, which compensate for the smoothing of the disc edge. 
For \object{DW UMa} with $i\sim82^\circ$, we may conclude that emission 
of the outer disc rim cannot be held responsible for the flat intensity profile, 
unless it is very bright or/and high. Note also that with increasing rim height
the scatter of the reconstructed intensities at given radius also increases.
The scatter in the maps of  \object{DW UMa} is nearly twice less than that
in the simulation with the rim height  $0.13R_{\rm disc}$, suggesting that the 
contribution of the rim in \object{DW UMa} is not very significant.
We conclude then that the flat temperature profile is not caused by 
the operation of the algorithm and the assumption of flat disc geometry.

Finally, we would like to note that the \object{SW Sex} stars are perhaps
the most complex of all types of CVs (Warner \cite{war}). The presence of 
overflowing stream, possibly a flared accretion disc, accretion disc wind and
a magnetic white dwarf violate some of the assumptions of eclipse mapping. 
In addition, many of the \object{SW Sex} stars show superhumps, both "positive"
and "negative". The former are now well understood to arise in an elliptic, 
progradely precessing accretion disc and will certainly complicate the 
situation ever more.  More generally, the presence of hot spots and
spiral shocks should also be 
considered. Detailed investigation of the effect of all these factors on
the eclipse mapping is a very complicated task and is out of the scope of this article. 
We will only  briefly discuss some of these topics.

The largest contribution of the spiral shocks should be $\sim10-15$\%
(e.g. Baptista et al. \cite{bap00,bap02}). When spread over
the disc, this will be much smaller and can hardly affect the radial profile. 
Pre-eclipse humps similar to those in quiescent dwarf novae are almost never 
observed in \object{SW Sex} stars, suggesting little contribution of the hot spot,
if any at all. This is expected though. The traces of the hot spot in the eclipses
of dwarf novae disappear in outburst. This generally implies little contribution 
of the spot when the disc is hot. Because the discs in \object{SW Sex} stars are hot,
the spot should little affect the eclipse mapping results
of these objects. Probably this is also the case with the superhumps, 
whose light source in NLs contributes no more than 10-20\% of the total system light.
A bigger problem with superhumps would be the different disc geometry, for example 
the elliptic disc shape during the "positive" superhumps. 
At least in one system, \object{PX And} 
(Stanishev et al. \cite{stpx}), the eclipse depth varies greatly with the "negative"
superhumps phase. This effect may be due to the presence of precessing tilted 
accretion disc and would have serious implications 
if the inclination is close to $90^\circ$.
Thus, both superhumps will affect the eclipse profile. However, when averaging many 
eclipses observed at different superhump phases, the effect of the superhumps will 
be reduced and  as a result average disc maps will be obtained. 
Other possible effects such as the  
spirals being thicker than the rest of the disc and any light scattered in the 
disc wind or/and the overflowing stream are difficult to take into account 
and estimate. In conclusion, we may say 
that various systematic effects may play a role in the eclipse mapping reconstructions
of the high-inclination NLs. Thus, the results of such experiments should be 
always discussed with this in mind.

\section{Summary}

The results of this study can be summarized as follows:

\begin{itemize}

\item Photometric observations obtained in intermediate and high states
show very large difference in the eclipse depth: $\sim1.2$ mag in
the high and $\sim3.4$ mag in the intermediate state. Eclipse mapping
reveals that this is due to variations of the accretion disc radius
resulting in the accretion disc being entirely eclipsed in the intermediate state,
whereas in the high state, a large part of the disc remains visible at
mid-eclipse. We found that the accretion disc radius increased from
$\sim0.5R_{\rm L_1}$ in the intermediate state to $\sim0.75R_{\rm
L_1}$ in the high state ($R_{\rm L_1}$ is the distance from the WD to
the first Lagrangian point $L_1$);

\item The central intensities of the high and intermediate state accretion disc
maps are very close. Thus, the increase of the disc radius fully
accounts for the final rise after the 1999/2000 low state.

\item The accretion disc brightness distribution is very flat. We suggest that
this might be due to the presence of a dense wind layer on top of
the accretion disc surface. This layer shades inner hot part of the disc and
the white dwarf. This simple model may account for: (1) the flat
disc profile; (2) the red UV spectrum in high state, and (3) nearly
the same temperature of the inner disc in high and intermediate
state.

\item Based on synthetic stellar spectra we investigate the
importance of limb-darkening when calculating the temperature of
optically thick accretion discs from eclipse maps. We find that the dependence of
limb-darkening on the temperature can have a significant effect on
the inferred disc temperatures, and thus should be always taken
into account.

\item  
The reliability of the reconstruction of 3D discs if a simple flat 1D disc 
geometry is used in the reconstruction is tested. We explored the case of a 
steady-state disc seen at inclination $i=82^\circ$ and possessing an emitting 
outer rim of different heights. The temperature of the rim is assumed 
equal to that of the outermost part of the disc. It is concluded that 
the radial profile of such a disc may still be well reconstructed 
assuming a flat geometry, provided that the height of the
disc rim is less than $\sim0.1R_{\rm disc}$.
If the disc rim is larger than that, the 1D reconstructed disc 
shows a larger scatter than is presently observed in \object{DW UMa}.
The importance of other effects that may introduce systematic errors
in the reconstruction of nearly edge-on discs is also discussed.

\item Periodogram analysis of the high state data reveals the
presence of "positive superhumps" with a period of $P_{\rm
sh}^+=0\fd1455$, in accord with the results of Patterson et al.
However, we cannot confirm the quasi-periodic oscillations
reported by these authors. In 2003 we also detect 
"positive superhumps" with a period of $P_{\rm sh}^+=0\fd1461$.

\item We obtain an updated orbital ephemeris:
$T_{\rm min}[HJD]=2446229.00687(9)+0\fd136606527(3)E$.

\end{itemize}

In conclusion we want to emphasize the importance of doing time
series photometry and spectroscopy of eclipsing \object{SW Sex}
stars which show low states and particularly during their low/high
state transitions. This may give us important clues for
understanding their nature and for accretion disc physics in
general.

\begin{acknowledgements}
HMJB and CP wish to thank Prof. Wilhelm Seggewiss and Prof. Klaus Reif 
for generously
allocating time at Hoher List. We thank Kent Honeycutt for kindly
providing us with his RoboScope photometry of DW UMa in machine readable
format. Thierry Pauwels is acknowledged for his help in collecting
the data. V. Stanishev acknowledges a grant from the Belgian Federal
Office for Scientific, Federal and Cultural Affairs (OSTC) in the
frame of the project "Multicolor photometry and astrometry of
double and multiple stars". The work was partially supported by
NFSR of Bulgaria with the project No.~715/97 and by the 
IAP P5/36 project of the Belgian Fed. State.
\end{acknowledgements}

\end{document}